\documentclass[runningheads]{llncs}
\usepackage[T1]{fontenc}
\usepackage{graphicx}
\begin{document}
\title{Integrating AIs With Body Tracking Technology for Human Behaviour Analysis: Challenges and Opportunities}
\titlerunning{Integrating AIs With Body Tracking for Human Behaviour Analysis}
%
\author{Adrien Coppens
\orcidID{0000-0002-2841-6708} \and
Valérie Maquil
\orcidID{0000-0002-0198-3729}} 
%
%
\institute{Luxembourg Institute of Science and Technology\\ 
\email{adrien.coppens@list.lu}\\
}
\maketitle              
\begin{abstract}

The automated analysis of human behaviour provides many opportunities for the creation of interactive systems and the post-experiment investigations for user studies.
Commodity depth cameras offer reasonable body tracking accuracy at a low price point, without the need for users to wear or hold any extra equipment.
The resulting systems typically perform body tracking through a dedicated machine learning model, but they can be enhanced with additional AI components providing extra capabilities.
This leads to opportunities but also challenges, for example regarding the orchestration of such AI components and the engineering of the resulting tracking pipeline. In this paper, we discuss these elements, based on our experience with the creation of a remote collaboration system across distant wall-sized displays, that we built using existing and readily available building blocks, including AI-based recognition models.

\end{abstract}
%
%


\section{Introduction}

Due to their large size and high resolution, Wall-Sized Displays (WSDs) are frequently utilised for collaborative group work, especially for activities that require participants to view various data sources and visualisations at the same time~\cite{jakobsen2014up,Langner2019}. In a shared space, discussions among co-workers often incorporate physical gestures and other forms of non-verbal communication. This dynamic enables them to effectively maintain awareness of one another's actions and co-ordinate their efforts efficiently during collaborative tasks.

\begin{figure}
    \centering
    \includegraphics[width=.9\textwidth]{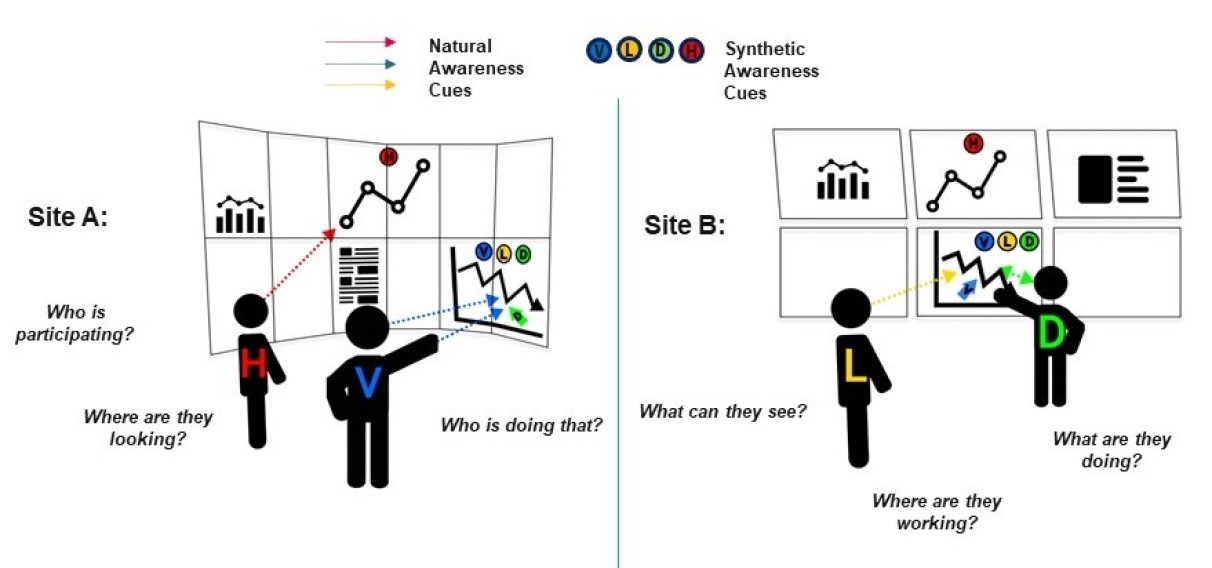}
    \caption{Workspace awareness for remote collaboration across Wall-Sized Displays. Reproduced from Anastasiou et al. \cite{anastasiou2024gesture}.}
    \label{fig:wa-diagram}
\end{figure}

In the context of the ReSurf project, we look at supporting remote collaboration across distant wall-sized displays. To do so, it is necessary to convey workspace awareness information~\cite{gutwin2002descriptive} on what local collaborators are doing to the remote site, to make up for the loss of natural awareness cues that collocated collaborators would gather~\cite{maquil2023establishing}. We therefore need to track the behaviour of local users, so that the corresponding information may be transmitted remotely through what we then refer to as synthetic awareness cues. This is illustrated through Figure \ref{fig:wa-diagram}.

Based on pilot observations and on comments collected during the early phases of the project, we opted to rely on the Azure Kinect sensor\footnote{\url{https://learn.microsoft.com/azure/kinect-dk/}} to drive the behaviour tracking system. To augment the resulting tracking capabilities with finer-grained gestures without introducing additional hardware devices, we wanted to explore the usage of image-based hand gesture recognition AI models. Similarly, in order to consistently assign authors to the information we transmit to the remote site, we explored image-based facial recognition.

This paper presents related work then covers some of the technical challenges we had to face, how we engineered our tracking pipeline based on the integration and orchestration of AI components, then discusses some remaining unexplored opportunities.

\section{Related Work}
\subsection{Behaviour analysis and action recognition}

Research has extensively explored interaction methods beyond the conventional mouse and keyboard, leading to interfaces that are sometimes termed post-WIMP~\cite{van1997post}. When it comes to WSDs, the need for alternative interfaces has become increasingly pressing, especially in order to limit the need for large mouse movements.

Touch-based systems are particularly common for WSDs, but that modality often requires physical navigation, which can be limiting in certain scenarios.
As a result, other solutions enabling interaction at a distance have been developed, such as mid-air pointing technologies~\cite{nancel2015mid}. 

To track these distant interactions, researchers have proposed various devices and techniques, including body-worn or hand-held objects. We refer to these as intrusive options because of the necessary extra equipment. Common examples are body suits for motion capture~\cite{kim2018deep}, telepointers (also known as wands~\cite{febretti2013cave2} or pens~\cite{fraser2007seconds}) used to perform pointing through raycasting~\cite{jota2010comparison}.

While these systems tend to offer greater precision and reliability, they are often cumbersome, requiring an initial setup step and potentially altering user behavior by providing a "tool" for interaction.

For this reason, we focus on non-intrusive technologies such as cameras. Standard RGB cameras can estimate body postures from two-dimensional images, enabling some level of interactivity. A notable library for this purpose is OpenPose~\cite{openpose}, which tracks full body movements, including hands and faces specifically~\cite{openpose-hand-face}.

However, standard cameras struggle with accurately inferring depth and distance from two-dimensional images, limiting their tracking capabilities. Although attempts have been made to use multiple cameras to improve depth estimation~\cite{chu2008real}, most non-intrusive tracking systems rely on depth cameras (RGB-D), which incorporate depth sensors alongside traditional RGB imaging. The Kinect family of depth cameras has reached wide usage in this context and has been extensively explored in prior research, due to its affordability and its reasonable performance. These sensors have been used for various applications, including video playback control~\cite{wittorf2016eliciting} and social network feed navigation~\cite{yoo2015dwell}. Some studies have focused on 3D interaction via two-dimensional virtual screens on which they may draw~\cite{kristensson2012continuous}, while others have focused on public displays, addressing how to engage passers-by~\cite{sousa2014deti}.

Researchers have also explored proximity-aware interactions~\cite{dostal2014spidereyes}, where the system displays information or activates input mechanisms~\cite{zhai2013gesture} based on how far the user stands from the large display. Furthermore, tracking has been combined with other modalities, such as speech and gaze input, to create seamless interaction experiences. The "Put-That-There" metaphor~\cite{bolt1980put} is a notable example of this, enabling the creation and manipulation of objects through a combination of pointing and speech.

While the studies mentioned cover various aspects of tracking for WSDs, focusing on local user interactions, there has been limited exploration of using such technologies to convey awareness information to a remote WSD.

\subsection{Awareness transmission across WSDs}
Some of the early approaches to transmit awareness information on WSDs aimed to create a sense of co-presence through full-screen video feeds~\cite{schremmer2007design,ishii1992clearboard,tang1991videowhiteboard}, similarly to videoconferencing solutions. However, such methods typically consume the entire display space, hindering data visualization and interaction.

Attempts to overcome this limitation include segmenting video feeds around users~\cite{kunz2010collaboard,kuechler2010collaboard} or adjusting their
transparency~\cite{edelmann2013preserving}, but these solutions struggle to scale effectively with additional participants and data elements.

The placement of smaller video feeds has also been explored and this option's capability to convey deictic references to remote participants has been evaluated~\cite{avellino2015accuracy}. However, such an approach relies on a specific known setup (display, camera, and feed placements).

Similarly to the cursors found in desktop-based collaborative systems, WSD-based solutions may rely on pointers to convey pointing gesture targets. These pointers should visually be distinct, to link each occurrence with its author's identity.

Robust user differentiation is crucial for assigning individual pointers and maintaining consistency even when users temporarily leave the tracking range. Previous methods using Kinect sensors placed behind users~\cite{wiechers2013user,von2016youtouch} faced challenges with accurate action attribution and gesture recognition.

Systems like YouTouch~\cite{von2016youtouch} and VTouch~\cite{chen2015vtouch} further worked on user identification through colour histograms and biometric measurements but remained sensitive to clothing changes, limiting long-term user recognition.

Advances in deep learning-based face recognition offer a promising solution. This technology can potentially overcome the limitations of previous approaches by enabling accurate and persistent user identification regardless of clothing variations. Integrating face recognition into the system could significantly enhance both usability and accuracy.



\section{Challenges}
Before diving into engineering aspects regarding the development of our tracking pipeline and its connection to AI models performing gestural and facial recognition, we briefly describe the challenges we encountered in constructing our RGB-D camera-based system.

\subsection{Scene calibration}

In order to transmit pointing and gazing cues, we need to retrieve the corresponding information using tracking technologies (in our case through RGB-D cameras). However, tracking devices will typically output data in their own coordinate system. In order to convert that data to screen-space coordinates, it is necessary to obtain a sufficiently accurate representation of the ``scene setup'' i.e. the dimensions and positions of the display with regards to the input camera. This for instance allows pointing and gazing information gathered from the tracking system to turn into pointing and gazing targets on screen.

The acquisition of such knowledge in itself typically involves a calibration step where human input is required.
This may involve the placement of markers at known locations so that they may be recognised by cameras~\cite{tuceryan1995calibration}
, or pointing-based solutions where users aim at keypoints (e.g. screen corners)~\cite{mills2021pointing}. These solutions might not always be particularly accurate, and some have instead opted to rely on fully manual measurements~\cite{gervais2015pointing}, which are themselves time-consuming.

To reduce the burden on the human calibrator but capitalise on the precise screen setup measurements we had access to (through an existing blueprint), we opted for an approach relying on adjustments based on point cloud streaming. We in fact stream the point cloud data coming from the camera to a Unity application that displays it over a virtual version of the scene setup, using precise screen measurements but approximate initial camera placement.

\begin{figure}
    \centering
    \includegraphics[width=.7\textwidth]{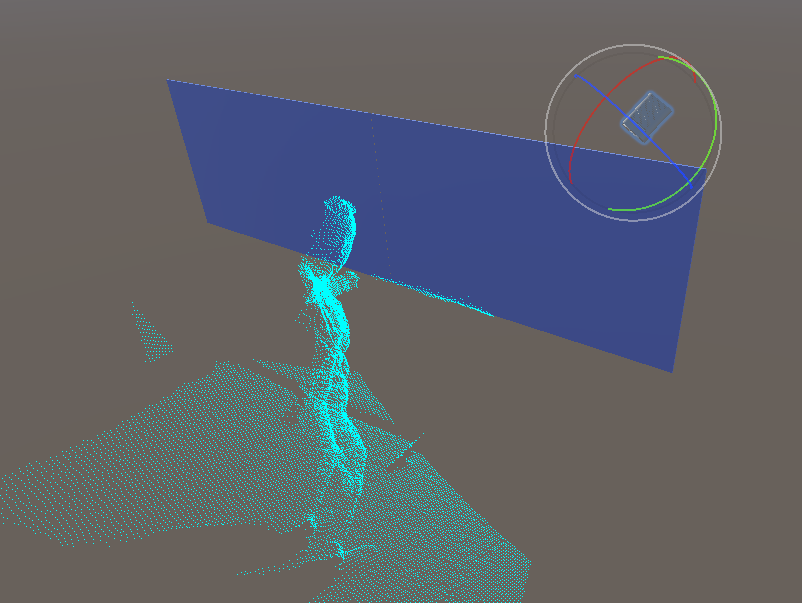}
    \caption{Scene calibration application in Unity, with two adjacent (dark blue) virtual screens whose dimensions correspond to their physical counterparts, a virtual camera, and a (cyan) point cloud created from the target camera's depth data. The placement of the camera can be adjusted so that the point cloud is aligned with the virtual screens so as to obtain a proper calibration matrix.}
    \label{fig:unity-calib}
\end{figure}

Figure \ref{fig:unity-calib} provides an example of the alignment process in that application. By adjusting the virtual camera's position and orientation so that the streamed point cloud matches the environment and the virtual screens are aligned, the precise placement of the camera in relation to the screens can be obtained. 

This approach works rather well and is sufficiently convenient for our case but scene calibration procedures based on (semi-)automated approaches using point cloud extraction~\cite{nguyen2013accurate} algorithms and possibly AI-based object recognition should be explored to make the process further easier while at least preserving the accuracy of the end result.

\subsection{Data fusion}

\begin{figure}
    \centering
    \includegraphics[width=.7\textwidth]{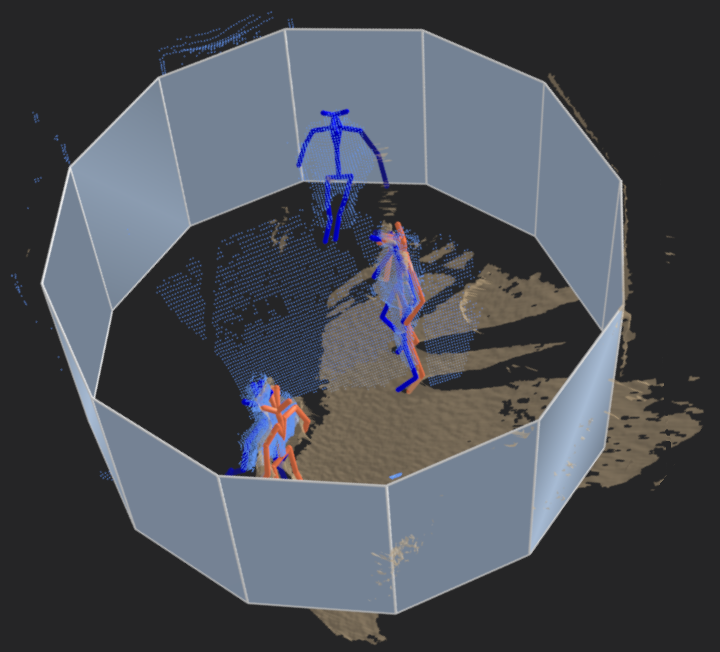}
    \caption{Point clouds and body tracking data from two cameras. The ``blue camera'' can see all three persons whereas the ``orange camera'' can only see two of them (the third person is currently occluded from that camera).}
    \label{fig:merging}
\end{figure}

A second challenging aspect relates to occlusion issues that tracking devices typically are subject to, as objects and users might get in the way and prevent the acquisition of suitable data for thereby temporarily ``hidden'' users. To alleviate that issue, we opted to rely on multiple cameras, as a person occluded from one camera might be visible for another one, as illustrated in Figure \ref{fig:merging}. This also increases the tracking range of the system as a whole.

However, combining data from multiple sensor requires the gathering of multi-sensor calibration information, typically through transformation matrices (one per additional camera, with each matrix converting data from its camera to the main camera i.e. the one whose position with regards to the screens is known). The approaches listed for the scene calibration step could be used to perform the multi-sensor calibration, by repeating the process for each sensor and using the scene calibration as the common frame of reference.
However, when cameras are placed in the same room, it is also possible to take advantage of common tracking areas. Indeed, if only one person is present in the room and two cameras are tracking that person, then the person can be used as a reference to determine the transformation matrix between the two cameras.

In an attempt to obtain more accurate calibration results, we have opted to additionally match the point clouds generated by these cameras, through the Iterative Closest Point~\cite{wang2017survey} algorithm and objects placed in the common tracking area (to generate ``common'' points to be matched). The resulting transformation matrix (between the point clouds) is also the transformation between the cameras.

Once obtained, such a transformation matrix can be used to convert body tracking data from one sensor's coordinate system to that of another sensor. There is however a chance that only one sensor see a given person (e.g. the blue skeleton on top of Figure~\ref{fig:merging}), and it may sometimes be difficult to identify which skeletons from different sensors belong to the same person. This leads to skeleton matching and merging problems as discussed in a separate paper~\cite{coppens2024skeletal}.

\section{Engineering an AI-based tracking pipeline}

\begin{figure}
    \centering
    \includegraphics[width=\textwidth]{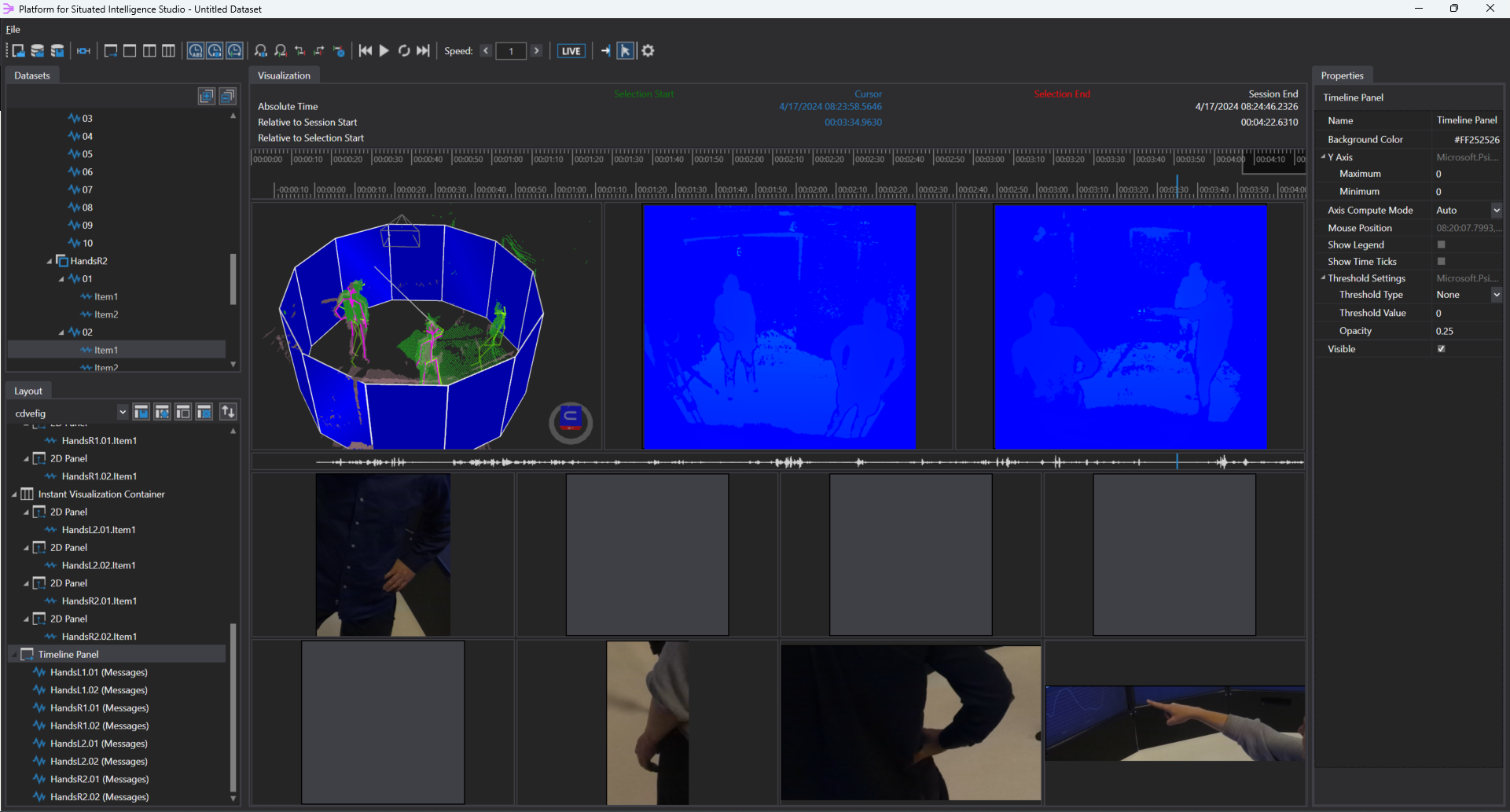}
    \caption{The graphical user interface of \textbackslash psi Studio, with an example layout displaying some of the data handled or created by our tracking pipeline.} 
    \label{fig:psi-studio}
\end{figure}

In order to build our tracking pipeline and combine the functionality of the body tracking, hand gesture recognition, and face recognition AIs, we rely on \textbackslash psi~\cite{bohus2021platform}. That framework facilitates the integration of AI components into multimodal systems by providing tools for managing, processing and combining streams of temporal data. It also greatly eases the development and debugging processes as well as the analysis of recorded data, by providing visualisation and data replaying tools (including \textbackslash psi Studio shown in Figure \ref{fig:psi-studio}).

As an example, two Azure Kinect sensors (with overlapping fields of view) cannot capture data at the same time; otherwise they will interfere with each other. This is solved by attributing different ``time slots" to each sensor, which is handled both by the Azure Kinect SDK\footnote{\url{https://github.com/microsoft/Azure-Kinect-Sensor-SDK}} and \textbackslash psi. However, the resulting data is therefore obviously captured at different times, and \textbackslash psi provides us with convenient interpolation features to counterbalance that issue, which helps improve the data fusion process through which a single stream of merged bodies is produced (for use by the rest of the pipeline).

The overall idea and structure of the tracking pipeline is that the data coming from the Azure Kinect sensor(s), including the aforementioned body tracking information, drives the system. Based on a known screen placement (relative to the camera), this generates pointing and gazing target points in screen-space coordinates.

\begin{figure}
    \centering
    \includegraphics[width=\textwidth]{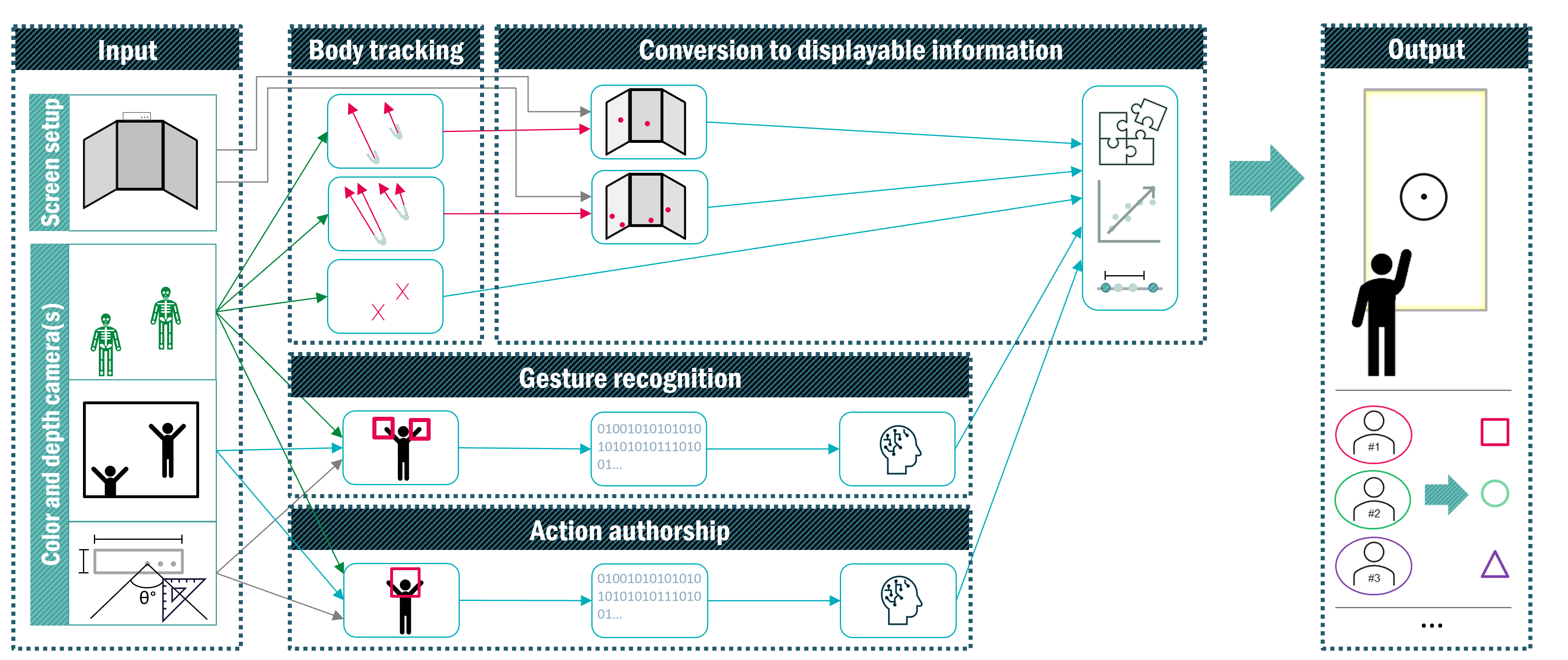}
    \caption{A high-level overview of the human behaviour tracking pipeline. Reproduced from Coppens et al.~\cite{coppens2024supporting}.} 
    \label{fig:pipeline}
\end{figure}

In addition to that, the same camera additionally provides color images. By combining the body tracking data and the camera's own intrinsics parameters, we can crop the color images around specific body parts (as seen in the bottom part of Figure \ref{fig:psi-studio}). This allow us to send the appropriate images to the AI components, respectively specialised in hand gesture and face recognition. This is done via messages in the MessagePack\footnote{\url{https://msgpack.org/}} format, transmitted via the ZeroMQ~\cite{rfczeromq} protocol.
The specialised recognition components can then feed their results back to the tracking pipeline, which aggregates the combined information to provide user-attributed behaviour information that includes pointing targets, gaze direction and hand gestures. Figure~\ref{fig:pipeline} presents a high-level overview of how that pipeline works. More details are available in the corresponding technical paper~\cite{coppens2024supporting}.

\section{Opportunities for embedding additional AIs} 
Outside of the opportunities linked to the enhancement and facilitation of the aforementioned scene and multi-camera calibration procedures, there are many opportunities for extending human behaviour analysis capabilities with other AI models.

Using a similar approach to the one we described here, one could in fact send the face crops to AI models trained to output information regarding gender~\cite{rouhsedaghat2021facehop}, age~\cite{agbo2021deep}, or emotions~\cite{li2020attention}.

Likewise, finer or different gestures could be supported through the integration of specialised AI models, and the presence of specific objects on or around users could be detected with dedicated AIs~\cite{rani2022three}.


Apart from image-based recognition models, automated human behaviour analysis systems may benefit from speech recognition AIs to transcribe what users are saying. Ideally, such transcriptions should be put in relation to precise timing information and attributed to specific speakers (e.g. to analyse turn-taking~\cite{wiemann2017turn}), through speaker diarisation AIs~\cite{park2022review}.

Interactive systems (possibly built on top of human behaviour analysis systems) may themselves benefit from speech synthesis AIs~\cite{kong2020hifi} to ``talk'' to system users. Combined with the recent advances in Large Language Models (LLMs)~\cite{zhao2023survey} that could be running in the background, these speech-based interfaces have the potential to produce compelling experiences. 
For example, Bohus et al. recently demonstrated~\cite{bohus2024sigma} the potential of combining scene understanding and object recognition AIs with LLMs and speech-based interaction for physical task assistance with a Hololens 2 headset\footnote{\url{https://www.microsoft.com/hololens}}.

We believe our tracking pipeline is a good basis to explore these kinds of opportunities for interactive systems based on unobtrusive human behaviour analysis in a room-scale context.
It can in fact adapt to hardware changes and software advances (e.g. replacement of an AI component by a better performing one) and can be extended to include additional AI models and capabilities.

\begin{credits}
\subsubsection{\ackname} Authors would like to thank the Luxembourg National Research Fund (FNR) for funding this research work under the FNR CORE ReSurf project (Grant nr C21/IS/15883550).

\subsubsection{\discintname}
The authors have no competing interests to declare that are
relevant to the content of this article.
\end{credits}
%
%
%
\bibliographystyle{splncs04}
\bibliography{sources.bib}

\begin{thebibliography}{10}
\providecommand{\url}[1]{\texttt{#1}}
\providecommand{\urlprefix}{URL }
\providecommand{\doi}[1]{https://doi.org/#1}

\bibitem{agbo2021deep}
Agbo-Ajala, O., Viriri, S.: Deep learning approach for facial age
  classification: a survey of the state-of-the-art. Artificial Intelligence
  Review  \textbf{54}(1),  179--213 (2021)

\bibitem{anastasiou2024gesture}
Anastasiou, D., Coppens, A., Maquil, V.: Gesture combinations during
  collaborative decision-making at wall displays. i-com (0) (2024)

\bibitem{avellino2015accuracy}
Avellino, I., Fleury, C., Beaudouin-Lafon, M.: Accuracy of deictic gestures to
  support telepresence on wall-sized displays. In: Proceedings of the 33rd
  annual acm conference on human factors in computing systems. pp. 2393--2396
  (2015). \doi{10.1145/2702123.2702448}

\bibitem{bohus2021platform}
Bohus, D., Andrist, S., Feniello, A., Saw, N., Jalobeanu, M., Sweeney, P.,
  Thompson, A.L., Horvitz, E.: Platform for situated intelligence (2021)

\bibitem{bohus2024sigma}
Bohus, D., Andrist, S., Saw, N., Paradiso, A., Chakraborty, I., Rad, M.: Sigma:
  An open-source interactive system for mixed-reality task assistance research
  - extended abstract. In: 2024 IEEE Conference on Virtual Reality and 3D User
  Interfaces Abstracts and Workshops (VRW). IEEE (2024)

\bibitem{bolt1980put}
Bolt, R.A.: “put-that-there” voice and gesture at the graphics interface.
  In: Proceedings of the 7th annual conference on Computer graphics and
  interactive techniques. pp. 262--270 (1980). \doi{10.1145/800250.807503}

\bibitem{openpose}
Cao, Z., Hidalgo, G., Simon, T., Wei, S.E., Sheikh, Y.: Openpose: Realtime
  multi-person 2d pose estimation using part affinity fields. IEEE Transactions
  on Pattern Analysis and Machine Intelligence  (2019).
  \doi{10.1109/TPAMI.2019.2929257}

\bibitem{chen2015vtouch}
Chen, Y., Liu, Z., Chou, P., Zhang, Z.: Vtouch: Vision-enhanced interaction for
  large touch displays. In: 2015 IEEE International Conference on Multimedia
  and Expo (ICME). pp.~1--6. IEEE (2015). \doi{10.1109/ICME.2015.7177390}

\bibitem{chu2008real}
Chu, C.W., Nevatia, R.: Real-time 3d body pose tracking from multiple 2d
  images. In: Articulated Motion and Deformable Objects. pp. 42--52. Springer
  (2008). \doi{10.1007/978-3-540-70517-8_5}

\bibitem{coppens2024supporting}
Coppens, A., Hermen, J., Schwartz, L., Moll, C., Maquil, V.: Supporting
  mixed-presence awareness across wall-sized displays using a tracking pipeline
  based on depth cameras. Proc. ACM Hum.-Comput. Interact.  \textbf{8}(EICS)
  (2024). \doi{10.1145/3664634}

\bibitem{coppens2024skeletal}
Coppens, A., Maquil, V.: Skeletal data matching and merging from multiple rgb-d
  sensors for room-scale human behaviour tracking. In: International Conference
  on Cooperative Design, Visualization and Engineering. pp. 289--298. Springer
  (2024). \doi{10.1007/978-3-031-71315-6_30}

\bibitem{dostal2014spidereyes}
Dostal, J., Hinrichs, U., Kristensson, P.O., Quigley, A.: Spidereyes: designing
  attention-and proximity-aware collaborative interfaces for wall-sized
  displays. In: Proceedings of the 19th international conference on Intelligent
  User Interfaces. pp. 143--152 (2014). \doi{10.1145/2557500.2557541}

\bibitem{edelmann2013preserving}
Edelmann, J., Mock, P., Schilling, A., Gerjets, P.: Preserving non-verbal
  features of face-to-face communication for remote collaboration",
  booktitle="cooperative design, visualization, and engineering. pp. 27--34.
  Springer Berlin Heidelberg (2013). \doi{10.1007/978-3-642-40840-3_4}

\bibitem{febretti2013cave2}
Febretti, A., Nishimoto, A., Thigpen, T., Talandis, J., Long, L., Pirtle, J.,
  Peterka, T., Verlo, A., Brown, M., Plepys, D., et~al.: Cave2: a hybrid
  reality environment for immersive simulation and information analysis. In:
  The Engineering Reality of Virtual Reality 2013. vol.~8649, pp. 9--20. SPIE
  (2013). \doi{10.1117/12.2005484}

\bibitem{fraser2007seconds}
Fraser, M., McCarthy, M.R., Shaukat, M., Smith, P.: Seconds matter: improving
  distributed coordination bytracking and visualizing display trajectories. In:
  Proceedings of the SIGCHI conference on Human Factors in Computing Systems.
  pp. 1303--1312 (2007). \doi{10.1145/1240624.1240822}

\bibitem{gervais2015pointing}
Gervais, R., Frey, J., Hachet, M.: Pointing in spatial augmented reality from
  2d pointing devices. In: Human-Computer Interaction--INTERACT 2015: 15th IFIP
  TC 13 International Conference, Bamberg, Germany, September 14-18, 2015,
  Proceedings, Part IV 15. pp. 381--389. Springer (2015)

\bibitem{gutwin2002descriptive}
Gutwin, C., Greenberg, S.: A descriptive framework of workspace awareness for
  real-time groupware. Computer Supported Cooperative Work (CSCW)  \textbf{11},
   411--446 (2002)

\bibitem{rfczeromq}
Hintjens, P., Hurton, M., Barber, I.: Zeromq message transport protocol. RFC 23
  (Jun 2020). \doi{10.17487/RFC9405}, \url{https://rfc.zeromq.org/spec/23/}

\bibitem{ishii1992clearboard}
Ishii, H., Kobayashi, M.: Clearboard: A seamless medium for shared drawing and
  conversation with eye contact. In: Proceedings of the SIGCHI conference on
  Human factors in computing systems. pp. 525--532 (1992).
  \doi{10.1145/142750.142977}

\bibitem{jakobsen2014up}
Jakobsen, M.R., Hornb{\ae}k, K.: Up close and personal: Collaborative work on a
  high-resolution multitouch wall display. ACM Transactions on Computer-Human
  Interaction (TOCHI)  \textbf{21}(2),  1--34 (2014). \doi{10.1145/2576099}

\bibitem{jota2010comparison}
Jota, R., Nacenta, M.A., Jorge, J.A., Carpendale, S., Greenberg, S.: A
  comparison of ray pointing techniques for very large displays. In: Graphics
  Interface. vol.~2010, pp. 269--276 (2010). \doi{10.5555/1839214.1839261}

\bibitem{kim2018deep}
Kim, D., Kwon, J., Han, S., Park, Y.L., Jo, S.: Deep full-body motion network
  for a soft wearable motion sensing suit. IEEE/ASME Transactions on
  Mechatronics  \textbf{24}(1),  56--66 (2018).
  \doi{10.1109/TMECH.2018.2874647}

\bibitem{kong2020hifi}
Kong, J., Kim, J., Bae, J.: Hifi-gan: Generative adversarial networks for
  efficient and high fidelity speech synthesis. Advances in neural information
  processing systems  \textbf{33},  17022--17033 (2020)

\bibitem{kristensson2012continuous}
Kristensson, P.O., Nicholson, T., Quigley, A.: Continuous recognition of
  one-handed and two-handed gestures using 3d full-body motion tracking
  sensors. In: Proceedings of the 2012 ACM international conference on
  Intelligent User Interfaces. pp. 89--92 (2012). \doi{10.1145/2166966.2166983}

\bibitem{kuechler2010collaboard}
Kuechler, M., Kunz, A.M.: Collaboard: a remote collaboration groupware device
  featuring an embodiment-enriched shared workspace. In: Proceedings of the
  2010 ACM International Conference on Supporting Group Work. pp. 211--214
  (2010). \doi{10.1145/1880071.1880107}

\bibitem{kunz2010collaboard}
Kunz, A., Nescher, T., Küchler, M.: Collaboard: A novel interactive electronic
  whiteboard for remote collaboration with people on content. In: 2010
  International Conference on Cyberworlds. pp. 430--437 (2010).
  \doi{10.1109/CW.2010.17}

\bibitem{Langner2019}
Langner, R., Kister, U., Dachselt, R.: {Multiple coordinated views at large
  displays for multiple users: Empirical findings on user behavior, movements,
  and distances}. IEEE Transactions on Visualization and Computer Graphics
  \textbf{25}(1),  608--618 (2019). \doi{10.1109/TVCG.2018.2865235}

\bibitem{li2020attention}
Li, J., Jin, K., Zhou, D., Kubota, N., Ju, Z.: Attention mechanism-based cnn
  for facial expression recognition. Neurocomputing  \textbf{411},  340--350
  (2020)

\bibitem{maquil2023establishing}
Maquil, V., Anastasiou, D., Afkari, H., Coppens, A., Hermen, J., Schwartz, L.:
  Establishing awareness through pointing gestures during collaborative
  decision-making in a wall-display environment. In: Extended Abstracts of the
  2023 CHI Conference on Human Factors in Computing Systems. pp.~1--7 (2023).
  \doi{10.1145/3544549.3585830}

\bibitem{mills2021pointing}
Mills, S.: Pointing-based calibration of a pointing interface for public
  displays. In: 2021 36th International Conference on Image and Vision
  Computing New Zealand (IVCNZ). pp.~1--6. IEEE (2021)

\bibitem{nancel2015mid}
Nancel, M., Pietriga, E., Chapuis, O., Beaudouin-Lafon, M.: Mid-air pointing on
  ultra-walls. ACM Transactions on Computer-Human Interaction (TOCHI)
  \textbf{22}(5),  1--62 (2015). \doi{10.1145/2766448}

\bibitem{nguyen2013accurate}
Nguyen, H.H., Kim, J., Lee, Y., Ahmed, N., Lee, S.: Accurate and fast
  extraction of planar surface patches from 3d point cloud. In: Proceedings of
  the 7th International Conference on Ubiquitous Information Management and
  Communication. pp.~1--8 (2013)

\bibitem{park2022review}
Park, T.J., Kanda, N., Dimitriadis, D., Han, K.J., Watanabe, S., Narayanan, S.:
  A review of speaker diarization: Recent advances with deep learning. Computer
  Speech \& Language  \textbf{72},  101317 (2022)

\bibitem{rani2022three}
Rani, S., Lakhwani, K., Kumar, S.: Three dimensional objects recognition \&
  pattern recognition technique; related challenges: A review. Multimedia Tools
  and Applications  \textbf{81}(12),  17303--17346 (2022)

\bibitem{rouhsedaghat2021facehop}
Rouhsedaghat, M., Wang, Y., Ge, X., Hu, S., You, S., Kuo, C.C.J.: Facehop: A
  light-weight low-resolution face gender classification method. In: Pattern
  Recognition. ICPR International Workshops and Challenges: Virtual Event,
  January 10-15, 2021, Proceedings, Part VIII. pp. 169--183. Springer (2021)

\bibitem{schremmer2007design}
Schremmer, C., Krumm-Heller, A., Vernik, R., Epps, J.: Design discussion of the
  [braccetto] research platform: Supporting distributed intensely collaborating
  creative teams of teams. In: International Conference on Human-Computer
  Interaction. pp. 722--734. Springer (2007).
  \doi{10.1007/978-3-540-73111-5_81}

\bibitem{openpose-hand-face}
Simon, T., Joo, H., Matthews, I., Sheikh, Y.: Hand keypoint detection in single
  images using multiview bootstrapping. In: IEEE Conference on Computer Vision
  and Pattern Recognition (CVPR) (2017)

\bibitem{sousa2014deti}
Sousa, T., Cardoso, I., Parracho, J., Dias, P., Sousa~Santos, B.:
  Deti-interact: interaction with large displays in public spaces using the
  kinect. In: Distributed, Ambient, and Pervasive Interactions: Second
  International Conference, DAPI 2014, Held as Part of HCI Interational 2014,
  Heraklion, Crete, Greece, June 22-27, 2014. Proceedings 2. pp. 196--206.
  Springer (2014). \doi{10.1007/978-3-319-07788-8_19}

\bibitem{tang1991videowhiteboard}
Tang, J.C., Minneman, S.: Videowhiteboard: video shadows to support remote
  collaboration. In: Proceedings of the SIGCHI conference on Human factors in
  computing systems. pp. 315--322 (1991). \doi{10.1145/108844.108932}

\bibitem{tuceryan1995calibration}
Tuceryan, M., Greer, D.S., Whitaker, R.T., Breen, D.E., Crampton, C., Rose, E.,
  Ahlers, K.H.: Calibration requirements and procedures for a monitor-based
  augmented reality system. IEEE Transactions on Visualization and Computer
  Graphics  \textbf{1}(3),  255--273 (1995)

\bibitem{van1997post}
Van~Dam, A.: Post-wimp user interfaces. Communications of the ACM
  \textbf{40}(2),  63--67 (1997). \doi{10.1145/253671.253708}

\bibitem{wang2017survey}
Wang, F., Zhao, Z.: A survey of iterative closest point algorithm. In: 2017
  Chinese Automation Congress (CAC). pp. 4395--4399. IEEE (2017)

\bibitem{wiechers2013user}
Wiechers, M., Nolte, A., Ksoll, M., Herrmann, T., Kienle, A.: User tracking for
  collaboration on interactive wall-sized displays. In: Mensch \& Computer. pp.
  191--200 (2013). \doi{10.1524/9783486781229}

\bibitem{wiemann2017turn}
Wiemann, J.M., Knapp, M.L.: Turn-taking in conversations. Communication theory
  pp. 226--245 (2017)

\bibitem{wittorf2016eliciting}
Wittorf, M.L., Jakobsen, M.R.: Eliciting mid-air gestures for wall-display
  interaction. In: Proceedings of the 9th Nordic Conference on Human-Computer
  Interaction. pp.~1--4 (2016). \doi{10.1145/2971485.2971503}

\bibitem{yoo2015dwell}
Yoo, S., Parker, C., Kay, J., Tomitsch, M.: To dwell or not to dwell: an
  evaluation of mid-air gestures for large information displays. In:
  Proceedings of the Annual Meeting of the Australian Special Interest Group
  for Computer Human Interaction. pp. 187--191 (2015).
  \doi{10.1145/2838739.2838819}

\bibitem{von2016youtouch}
von Zadow, U., Reipschl{\"a}ger, P., B{\"o}sel, D., Sellent, A., Dachselt, R.:
  Youtouch! low-cost user identification at an interactive display wall. In:
  Proceedings of the International Working Conference on Advanced Visual
  Interfaces. pp. 144--151 (2016). \doi{10.1145/2909132.2909258}

\bibitem{zhai2013gesture}
Zhai, Y., Zhao, G., Alatalo, T., Heikkil{\"a}, J., Ojala, T., Huang, X.:
  Gesture interaction for wall-sized touchscreen display. In: Proceedings of
  the 2013 ACM conference on Pervasive and ubiquitous computing adjunct
  publication. pp. 175--178 (2013). \doi{10.1145/2494091.2494148}

\bibitem{zhao2023survey}
Zhao, W.X., Zhou, K., Li, J., Tang, T., Wang, X., Hou, Y., Min, Y., Zhang, B.,
  Zhang, J., Dong, Z., et~al.: A survey of large language models. arXiv
  preprint arXiv:2303.18223  (2023)

\end{thebibliography}
%




\end{document}